\documentclass[
 reprint,
superscriptaddress,
prl,
 amsmath,amssymb,
 aps,
floatfix,
]{revtex4-1}

\usepackage{graphicx}
\usepackage{dcolumn}
\usepackage{bm}

\usepackage[mathlines]{lineno}
\usepackage{times}
\usepackage{fouriernc}
\usepackage{bookmark}
\usepackage[colorinlistoftodos]{todonotes}
\usepackage[latin1]{inputenc}
\usepackage{hyperref}

\newcommand{\lacu}{LaCu$_{6-x}$Au$_{x}$~}

\begin{document}

\preprint{APS/123-QED}

\title{Candidate Elastic Quantum Critical Point in LaCu$_{6-x}$Au$_x$}

\author{L. Poudel}
\email{lpoudel@vols.utk.edu}
\affiliation{Department of Physics \& Astronomy, University of Tennessee, Knoxville, TN-37966, USA}
\affiliation{Quantum Condensed Matter Division, Oak Ridge National Laboratory, Oak Ridge, TN-37831, USA}

\author{A. F. May}%
\affiliation{Materials Science \& Technology Division, Oak Ridge National Laboratory, Oak Ridge, TN-37831, USA
}

\author{M. R. Koehler}
\affiliation{Department of Material Science \& Engineering, University of Tennessee, Knoxville, TN-37966, USA}

\author{M. A. McGuire}
\affiliation{Materials Science \& Technology Division, Oak Ridge National Laboratory, Oak Ridge, TN-37831, USA
}

\author{S. Mukhopadhyay}
\affiliation{Materials Science \& Technology Division, Oak Ridge National Laboratory, Oak Ridge, TN-37831, USA
} 
\author{S. Calder}
\affiliation{Quantum Condensed Matter Division, Oak Ridge National Laboratory, Oak Ridge, TN-37831, USA}

\author{R. E. Baumbach}
\affiliation{National High Magnetic Field Laboratory, Florida State University, Tallahassee, Florida 32306, USA}

\author{R. Mukherjee}
\affiliation{Department of Material Science \& Engineering, University of Tennessee, Knoxville, TN-37966, USA}
 
\author{D. Sapkota}
\affiliation{Department of Physics \& Astronomy, University of Tennessee, Knoxville, TN-37966, USA}
 
\author{C. de la Cruz}
\affiliation{Quantum Condensed Matter Division, Oak Ridge National Laboratory, Oak Ridge, TN-37831, USA}

\author{D. J. Singh}
\affiliation{Department of Physics \& Astronomy, University of Missouri, Columbia, MO-65211, USA}
 
\author{D. Mandrus}
\affiliation{Department of Physics \& Astronomy, University of Tennessee, Knoxville, TN-37966, USA}
\affiliation{Department of Material Science \& Engineering, University of Tennessee, Knoxville, TN-37966, USA}
\affiliation{Materials Science \& Technology Division, Oak Ridge National Laboratory, Oak Ridge, TN-37831, USA
}

\author{A. D. Christianson}
\affiliation{Quantum Condensed Matter Division, Oak Ridge National Laboratory, Oak Ridge, TN-37831, USA} 
\affiliation{Department of Physics \& Astronomy, University of Tennessee, Knoxville, TN-37966, USA}

\date{\today}

\begin{abstract}
The structural properties of LaCu$_{6-x}$Au$_x$ have been studied using neutron diffraction, x-ray diffraction, and heat capacity measurements. The continuous orthorhombic-monoclinic structural phase transition in LaCu$_{6}$ is suppressed linearly with Au substitution until a complete suppression of the structural phase transition occurs at the critical composition, $x_{c}$ = 0.3. Heat capacity measurements at low temperatures indicate residual structural instability at $x_c$ that extends well into the orthorhombic phase. The instability is ferroelastic in nature, with density functional theory (DFT) calculations showing negligible coupling to electronic states near the Fermi level. The data and calculations presented here are consistent with the zero temperature termination of a continuous structural phase transition suggesting that the LaCu$_{6-x}$Au$_x$ series hosts an elastic quantum critical point. 

\end{abstract}

\maketitle

Quantum fluctuations in the vicinity of a quantum critical point (QCP) generate a seemingly inexhaustible supply of new phases of matter \cite{Park2006,Nakatsuji2008,stewart,RevModPhys.79.1015, Mathur1998,Sachdev475,Lake,keknzelmann,ruegg}. Although of great interest, the multiple competing phases typically found near a QCP obscure the fundamental critical behavior and consequently a general understanding of such phase transitions has remained elusive and no concept as powerful as the universality of classical continuous phase transitions has emerged. Hence, new examples of QCPs are highly prized as a means of probing the general organizing principles of QCPs. Thus it is rather surprising, given the immense body of work devoted to the study of structural phase transitions at finite temperature, that little attention has been given to the concept of a structural quantum critical point (SQCP). The bulk of previous work on QCPs involving structural degrees of freedom has been focused on an incipient ferroelectric state (\textit{e.g.} Refs. \cite{oppermann,folk,schneider,millev,scott,morf,Rowley2014}.) For example, the quantum zero point motion in SrTiO$_3$ acts to prevent the complete softening of the relevant optic phonon mode resulting in a quantum paraelectric on the verge of ferroelectricity \cite{srtio3,sttio3_pressure}. 

Recently, interest in SQCPs has risen. For instance, ScF$_3$ has been discussed as an example of an SQCP in an ionic insulator \cite{scf3}. On the other hand, in metallic systems, there have been several studies of a family of Rameika phase stannides where a structural phase transition with concomitant changes in the Fermi surface can be tuned to an SQCP \cite{ambient_goh,strong_coupling,SQCP_IR,strongenh_biswas,lattice_kuo,kuo_Sr3Ir4Sn13,Fang}. Here, we approach the problem of an SQCP from the perspective of a structural phase transition where elastic instabilities are primarily responsible for the lowering of symmetry and electronic degrees of freedom are unimportant. This type of transition  is similar to the nematic transition, as found, for example in the Fe-based superconductors \cite{fang_nematic,xu,Fradkin155}. The similarity being that the point group symmetry is lowered without a change in translational symmetry. The two cases are nevertheless distinct, as in the Fe-based superconductors, the nematic transition is strongly coupled to electronic states near the Fermi level, whereas the elastic transition studied here is in the opposite decoupled limit.


Elastic QCPs, where the quantum zero point motion of the atoms destroys an elastically ordered state, have been the subject of recent theoretical study \cite{zacharias}. However, examples where electronic degrees of freedom are uninvolved have, to the best of our knowledge, eluded study thus far. To remedy this, we have identified the LaCu$_{6-x}$Au$_x$ series as a promising candidate to host an elastic QCP. LaCu$_6$ exhibits a continuous phase transition from an orthorhombic ($Pnma$) to a monoclinic ($P2_1/c$) crystal structure at 460 K \cite{yamada}. A phase transition of this type is a common feature of the $R$Cu$_6$ ($R$ = La, Ce, Pr, Nd, Sm) family including the CeCu$_{6-x}$Au$_x$ series which hosts one of the most studied antiferromagnetic QCPs \cite{PrCu6,RCu6,rus_cecu6,yamada,grube1999suppression, schroder2000onset, schroder1998scaling,stockert,RevModPhys.79.1015}. The continuous nature of the orthorhombic-monoclinic phase transition is evidenced by: a smooth evolution of the monoclinic distortion \cite{PrCu6,RCu6}; a gradual softening of the $C_{66}$ elastic constant \cite{rus_cecu6}; a progressive softening of a transverse acoustic phonon mode \cite{yamada}; and a continuous change in the linear thermal expansion coefficient \cite{grube1999suppression}. Furthermore, no signature of the structural phase transition is visible in resistivity measurements \cite{transport}, suggesting that the electronic degrees of freedom are of little consequence. Finally, among the different members of the $R$Cu$_6$ family, the absence of unpaired $f-$electrons in LaCu$_6$ eliminates complications due to a magnetic ground state. 

In this letter, we present a detailed study of the structural properties of the LaCu$_{6-x}$Au$_x$ series as a function of Au-composition. The structural transition temperature ($T_S$) in LaCu$_{6-x}$Au$_x$ decreases linearly with Au-doping, and extrapolation of $T_S$ indicates complete suppression of the transition at the critical composition $x_c$ = 0.3. The low temperature heat capacity is maximum at $x_c$ as is typical for a QCP. DFT calculations show that the Fermi surface is essentially unchanged at $T_S$ and moreover that the structural transition occurs as the result of a zone center elastic instability. The measurements taken together with the calculations point towards an elastic QCP in the \lacu series.

\begin{figure}[t]
\centering
\includegraphics[width=0.45\textwidth]{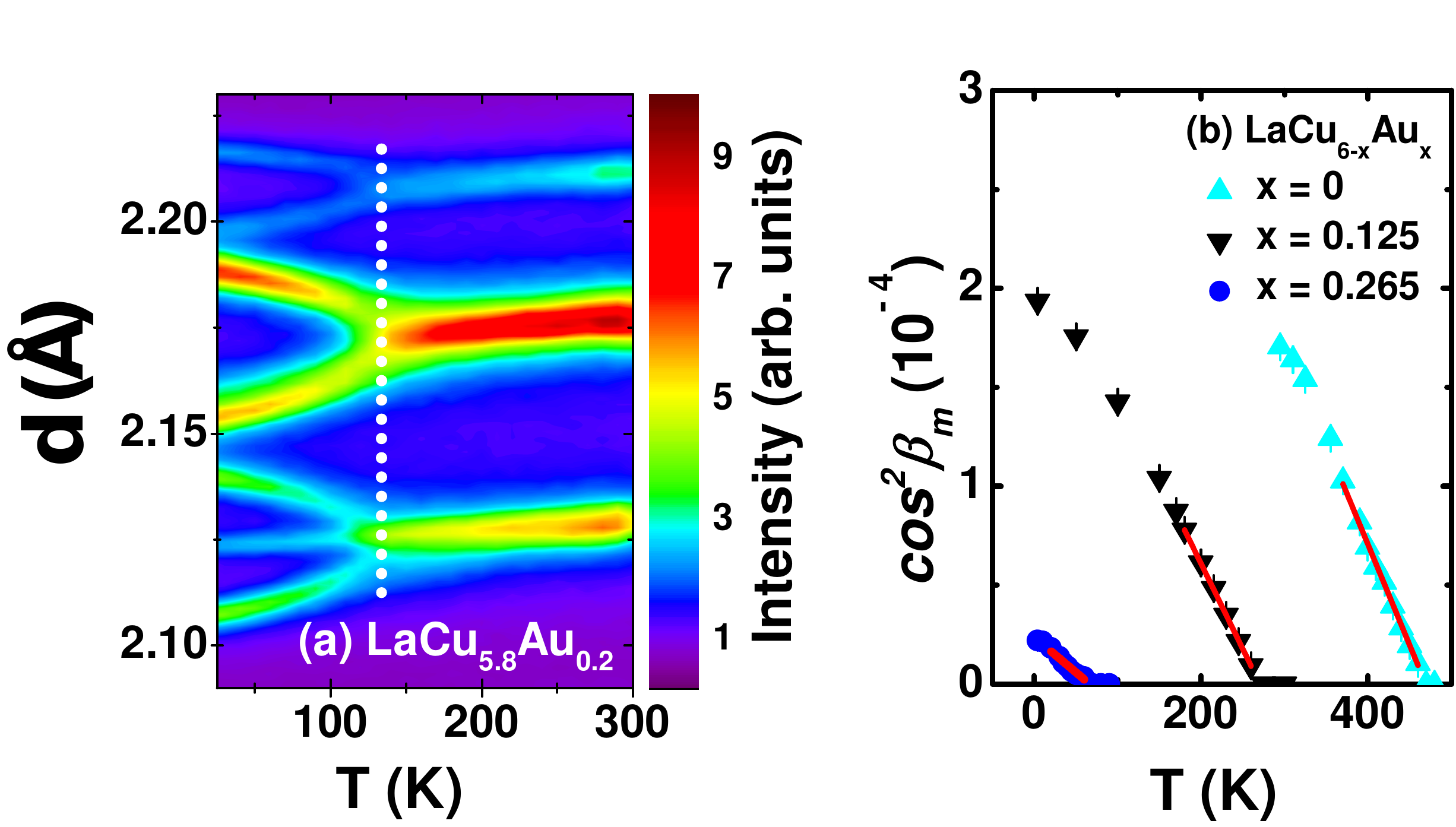}
\caption{(a) X-ray diffraction data showing the splitting of selected Bragg peaks at the structural transition in LaCu$_{5.8}$Au$_{0.2}$. The following indexing scheme describes the splitting of the peaks: (122)$\rightarrow$(22$\pm$1), (220)$\rightarrow$(20$\pm$2), (221)$\rightarrow$(21$\pm$2). The peak between (21$\pm$2) at low temperature is the orthorhombic (303) reflection, which becomes the monoclinic (033) reflection (no splitting). The vertical dotted line indicates the temperature where the peaks begin to separate. Note: Each pixel represents a triangulation-based linear interpolation of the measurement performed with 10 K steps in temperature. (b)  The temperature dependence of $\cos^2{\beta_m}$ in LaCu$_{6-x}$Au$_{x}$ determined from neutron diffraction data. $\cos^2{\beta_m}$ varies smoothly with temperature near $T_S$. For each composition, $T_S$ was estimated from a linear extrapolation (red line) of $\cos^2{\beta_m}$.
}
\label{st_transition}
\end{figure}


Polycrystalline samples of LaCu$_{6-x}$Au$_x$ were prepared by arc melting stoichiometric compositions of La (purity $\geq$ 99.998\%, Ames laboratory), Cu (purity 99.9999\%, Alpha Aesar) and Au (purity 99.9999 \%, Alpha Aesar). Specific heat measurements were performed on selected compositions. Structural properties for each composition were obtained from Rietveld analysis of neutron and/or x-ray diffraction patterns. Further details of sample characterization are provided in supplementary material \cite{supp}. The lattice parameters transform from orthorhombic to monoclinic symmetry as follows: $a$ $\rightarrow$ $c_m$, $b$ $\rightarrow$ $a_m$, $c$ $\rightarrow$ $b_m$, and $\beta_m$ denotes the monoclinic angle.


The structural phase transition in LaCu$_6$ was tuned by substituting Cu by isoelectronic Au. The substitution results in an overall expansion of the unit cell due to the larger atomic radius of Au (covalent radius = 144 fm) compared to Cu (covalent radius = 128 fm), which is typical for the $R$Cu$_6$ family \cite{grube1999suppression}. With Au substitution for ($x\leq0.7$), the lattice parameters $a$ and $c$ increase whereas $b$ decreases \cite{supp}. For each composition studied, the lattice parameters increase smoothly with temperature and no discontinuities were observed at $T_S$ \cite{supp}.

To examine the impact of Au-doping in greater detail, high resolution synchrotron x-ray measurements were performed on two compositions: LaCu$_6$ (monoclinic) and LaCu$_{5.7}$Au$_{0.3}$ (orthorhombic) \cite{supp}. In the orthorhombic ($Pnma$) structure, the La occupies a $4c$ site, whereas the Cu atoms are distributed among one $8d$ and four $4c$ sites. The Au atoms in LaCu$_{5.7}$Au$_{0.3}$ occupy only a particular $4c$ site: Cu2 (0.146,~0.25,~0.139). The same site preference was also observed in the related materials CeCu$_{6-x}$Au$_x$ \cite{Ruck:se0121} and CeCu$_{6-x}$Ag$_x$ \cite{poudel} and is attributed to the large volume of the Cu2 site \cite{grube1999suppression}. At the structural transition, the symmetry of all $4c$ sites remains unchanged, whereas the orthorhombic $8d$ site separates into two monoclinic $4c$ sites due to the loss of the mirror symmetry along the (100) plane.

\begin{figure}[t]
\centering
\includegraphics[width=0.45\textwidth]{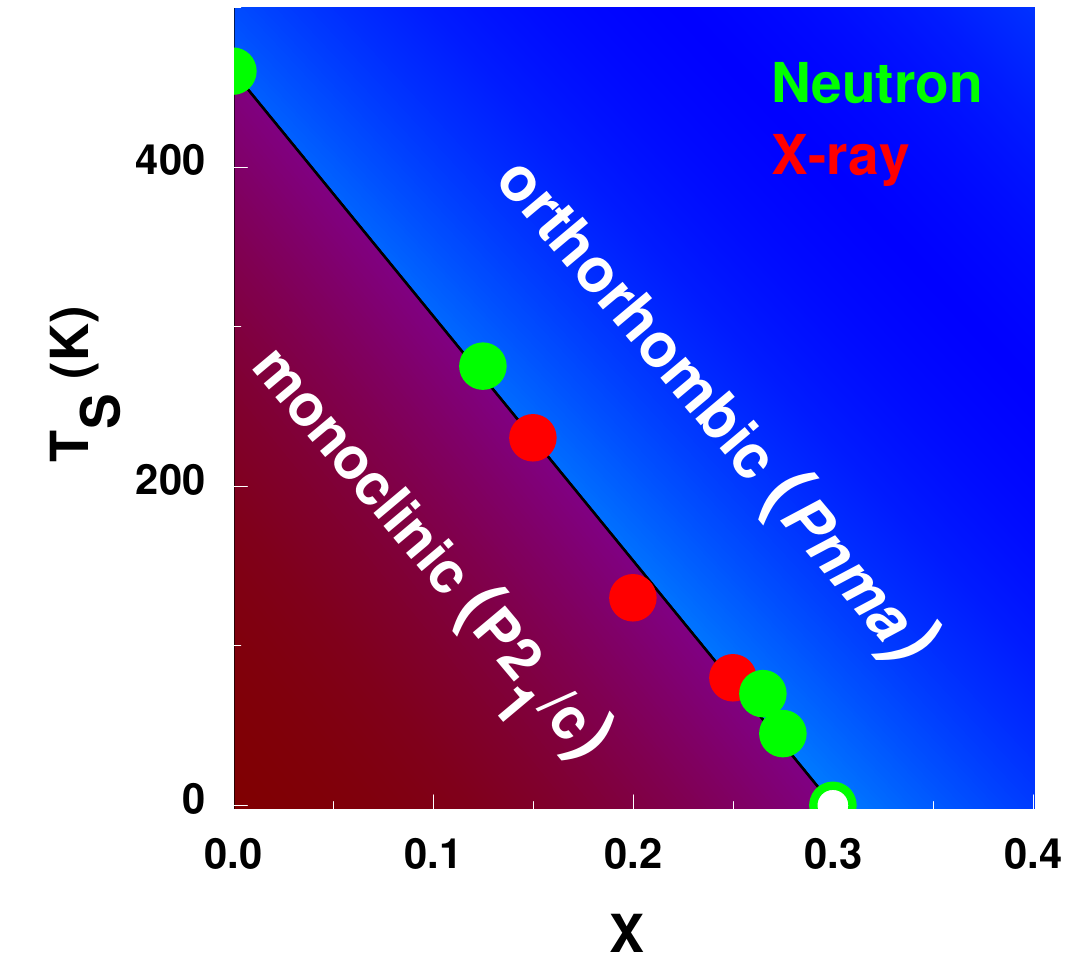}
\caption{Structural phase diagram of LaCu$_{6-x}$Au$_{x}$. The open circle represents the point where no structural phase transition is observed above 2 K. The statistical error bars for $T_S$ are smaller than the symbol size. Linear extrapolation of $T_S$ with temperature indicates $x_c~=~0.3$. }
\label{Auphase_hori}
\end{figure}

The monoclinic distortion results in the splitting of Bragg peaks as shown in Fig. \ref{st_transition}(a). The shear strain $e_{13} = \frac{1}{2}\frac{c}{c_o}  \cos\beta_m$ \cite{carpenter1998} acts as an order parameter for the phase transition. Hence to determine $T_S$ for each composition, the values of $\cos^2{\beta_m}$ extracted from the analysis of the diffraction data were extrapolated as a function of temperature (Fig. \ref{st_transition}(b)). The resulting transition temperatures in \lacu are summarized in the phase diagram shown in Fig. \ref{Auphase_hori}. A linear decrease in $T_S$ with Au-composition is observed. Extrapolation of $T_S$ as a function of Au-composition shows that the phase transition disappears at $x_{c}$ = 0.30(3) \cite{note1}. Further support for this value of $x_c$ and the presence of a QCP is provided by the extrapolation of $\cos^2{\beta_m}$ at 20 K for each composition, which yields $x_c$ = 0.28 \cite{supp}.

\begin{figure}[t]
\centering
\includegraphics[width=0.45\textwidth]{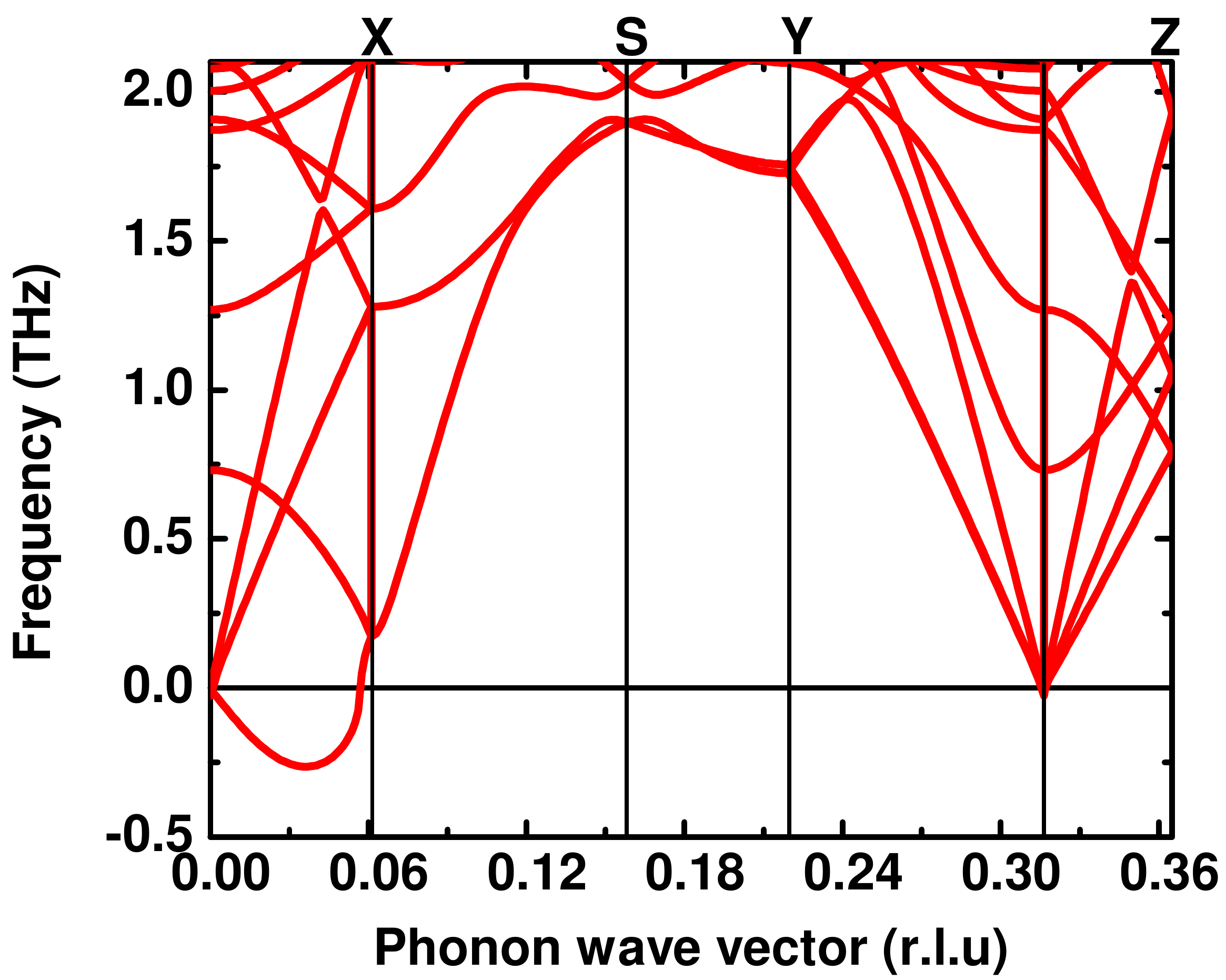}
\caption{Calculated phonon dispersions of LaCu$_6$ with the $Pnma$ crystal structure. The imaginary phonon mode frequency from $\Gamma$ to $X$ indicates the instability of the orthorhombic structure. }
\label{phonon}
\end{figure}

First principles calculations provide several additional insights into the nature of the structural phase transition in LaCu$_{6-x}$Au$_x$. DFT calculations using the Perdew, Burke and Enzerhof (PBE) generalized gradient approximation \cite{perdew} were performed for LaCu$_6$ with experimental structures. Phonon calculations were performed with the Phonopy code \cite{phonopy} based on underlying DFT calculations by the VASP code \cite{blochl,kresse}. Electronic structure calculations were performed with the general potential linearized augmented planewave (LAPW) method \cite{kresse} as implemented in WIEN2k \cite{singh,wien}. The calculations were done using well converged basis sets. LAPW plus local orbital basis sets were used with LAPW sphere radii of 2.3 $bohr$ for La and 2.25 $bohr$ for Cu. The electronic structure calculations were performed using experimental lattice parameters for the orthorhombic and monoclinic cells with fully relaxed internal atomic coordinates.

The phonon dispersions for the high temperature $Pnma$ structure (Fig. \ref{phonon}) show a clear instability of ferroelastic character. In particular, an unstable transverse acoustic branch is found along $\Gamma-X$, and this branch has an upward curvature, meaning that the instability is a zone center elastic instability. There are no soft optic phonons. Such zone center symmetry lowering without other apparent long range orders can reflect nematicity, associated with orbital or magnetic degrees of freedom as has been suggested in the Fe-based superconductors, for example \cite{fang_nematic,xu,Fradkin155}, Pomeranchuk instabilities of the Fermi surface \cite{Pomeranchuk}, or lattice instability due to steric effects, $i.e.$ under-bonding or poor matches between ion sizes and the lattice structure. Our electronic structure calculations indicate that the latter is the case in LaCu$_6$. The phonon softening calculated by DFT is in line with the softening of the transverse acoustic mode observed in CeCu$_6$ and LaCu$_6$ \cite{yamada}. The transition is continuous and originates from a zone center elastic instability, and may thus be termed ferroelastic, although we have not observed switching in the low-T phase and therefore formally it is coelastic \cite{salje}. Further investigations to confirm ferroelasticity in \lacu are essential.  

The electronic densities of states are very similar for the monoclinic and orthorhombic phases. In particular, there is no significant change around the Fermi level, E$_F$ \cite{supp}. Similar to the prior calculations \cite{harima}, the Cu $d$-bands are occupied and are located from approximately -5 $eV$ to -1.7 $eV$ with respect to $E_F$ and therefore do not play a role in the low energy physics, which instead depends on an electronic structure derived mainly from Cu and La $s-$states. The Fermi surface is derived from five bands that cross $E_F$ and is similar in both two phases \cite{supp}. The calculated plasma frequencies($\Omega_p$) for the orthorhombic phase are  3.00 $eV$, 3.03 $eV$ and 3.61 $eV$ along $a$, $b$ and $c$, respectively. The Fermi level density of states is very similar between the two phases: $N(E_F)$ = 3.52 $eV^{-1}$ and 3.37 $eV^{-1}$ per formula unit for the orthorhombic and monoclinic phases, respectively.

\begin{figure*}[t]
\centering
\includegraphics[width=0.9\textwidth]{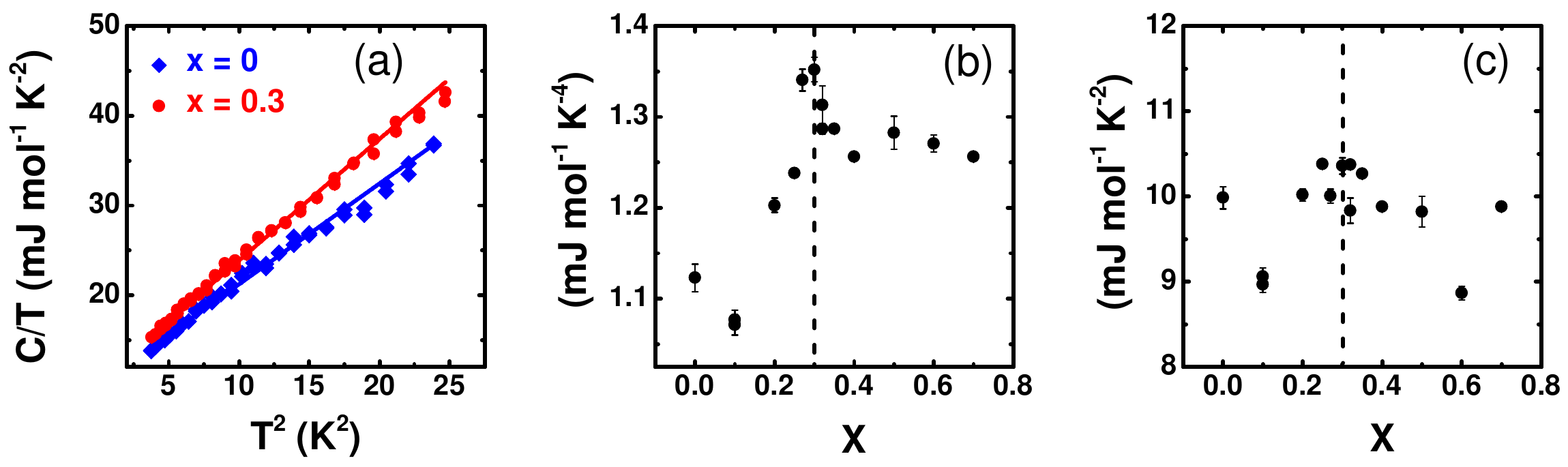}
\caption{(a) Heat capacity data showing $T^3$ behavior at low temperature for LaCu$_6$ ($x=0$) and LaCu$_{5.7}$Au$_{0.3}$ ($x_c$=$0.3$). The solid lines are a fit to the data of the form $C~=~\gamma T + \beta T^3$. (b,c) The variation of (b) $\beta$ and (c) $\gamma$ obtained from the fits to the heat capacity measurements as a function of Au-composition. The vertical dotted lines indicate the critical composition ($x_c$ = 0.3). In all panels, the values are reported per formula unit.
 }
\label{heat_capacity}
\end{figure*}

Thus the electronic structure of orthorhombic and monoclinic LaCu$_6$ are very similar to each other in all respects, including the electronic structure near $E_F$. This implies that the phase transition is not driven electronically, $e.g.$ by an instability of the Fermi surface such as a density wave, as there is no substantial reconstruction of the electronic structure associated with the transition. It also implies that the structural transition does not couple strongly to electrons at the Fermi energy, meaning that only weak effects in transport and other electronic properties may be expected. Together the unstable transverse acoustic branch along $\Gamma-X$ and the decoupled electronic degrees of freedom are consistent with a ferroelastic ground state in LaCu$_6$. We note that the variations of the transition temperature upon substitution on the La and Cu site are consistent with the characterization of the transition as a ferroelastic transition driven by size effects, specifically, substitution of a larger atom on the Cu site behaves similar to the substitution of a smaller atom on the La site \cite{grube1999suppression}, making a connection between the instability and size mismatch, as opposed to chemical pressure, band-width, $etc$.

Having determined that the structural transition is elastic in origin and that the electronic degrees of freedom are unimportant, we now examine the thermodynamic properties with heat capacity measurements. Typically at a QCP, the energy scale associated with the critical fluctuations vanishes resulting in an enhanced heat capacity as the QCP is approached--the softening of the transverse acoustic mode discussed above plays this role here. As anticipated, a larger heat capacity is observed at the putative QCP in \lacu (Fig. \ref{heat_capacity}(a)). The relation, $C~=~\gamma T + \beta T^3$ was utilized to parametrize the low temperature heat capacity for each composition, where the coefficients, $\gamma$ and $\beta$, represent the electronic and phonon contributions, respectively \cite{kittel}. As shown in the Fig. \ref{heat_capacity}(b,c), $\beta$ reaches a maximum value at $x_c$, while $\gamma$ remains essentially independent of Au-composition. The modest increase in the phonon heat capacity at $x_c$ is in line with the softening of the acoustic mode along $\Gamma-X$. The softening of a single acoustic mode is a small contribution to $\beta$ which contains contributions from all thermally accessible phonon modes. Interestingly, $\beta$ is elevated for compositions larger than $x_c$, that is the orthorhombic region of the phase diagram directly adjacent to the QCP. Measurements which directly probe the unstable acoustic mode would be particularly useful to further illuminate this issue. 

Recently, the (Ca$_x$Sr$_{1-x}$)$_3$Rh$_4$Sn$_{13}$ \cite{ambient_goh,strong_coupling} and the (Ca$_x$Sr$_{1-x}$)$_3$Ir$_4$Sn$_{13}$ \cite{SQCP_IR} series have been identified as systems that can be tuned to a displacive SQCP by application of physical or chemical pressure. In an intriguing analogy with many unconventional superconductors,  the phase diagrams of these systems comprise a dome of superconductivity peaked near an SQCP \cite{strong_coupling,strongenh_biswas,tompsett}. Furthermore, the structural phase transition in these systems occurs with an abrupt change in the Fermi surface associated with a charge density wave \cite{kuo_Sr3Ir4Sn13,strongenh_biswas,lattice_kuo,Fang,SQCP_IR}. Hence, there is a clear distinction with the QCP in LaCu$_{6-x}$Au$_x$ where there is no expectation (or evidence) that electronic degrees of freedom play an important role or are affected in a significant way. Also, an elastic QCP is associated with an instability of a acoustic mode rather than an optic mode as in the case of displacive SQCP. Nevertheless, there are some similarities, for example, the phonon specific heat is enhanced near the SQCP in LaCu$_{6-x}$Au$_x$ and somewhat more dramatically in the stannides mentioned above \cite{ambient_goh,SQCP_IR}. These observations indicate that the elastic QCP studied here is fundamentally different from the displacive SQCP and the \lacu series appears to be an ideal candidate system to further probe an elastic QCP without the complications of competing electronic phases.


%

In conclusion, we have studied the LaCu$_{6-x}$Au$_x$ system, where a continuous structural phase transition can be suppressed via Au substitution. A phase diagram showing that the complete suppression of the structural phase transition occurs for $x_c$ = 0.3 was constructed from the results of x-ray and neutron diffraction measurements. The low-temperature phonon contribution to the heat capacity rises at the critical composition, indicating residual instability of the orthorhombic structure at low temperature. These observations indicate that the suppression of the monoclinic phase with Au-substitution in LaCu$_{6-x}$Au$_x$ likely results in a QCP. DFT calculations support the idea that the QCP in LaCu$_{6-x}$Au$_x$ arises from an elastic instability with no significant involvement of electronic degrees of freedom. Further investigation of \lacu should prove invaluable in the elucidation of the fundamental properties of an elastic QCP.

\begin{acknowledgments}

We acknowledge  V. Keppens, V. Fanelli, T. Williams, and P. Whitfield for useful discussions, F. Ye for help with sample characterization, and M. Suchomel for assistance with the synchrotron x-ray measurements. The research at the High Flux Isotope Reactor (ORNL) is supported by the Scientific User Facilities Division, Office of Basic Energy Sciences, U.S. Department of Energy (DOE). Use of the Advanced Photon Source at Argonne National Laboratory was supported by the U. S. Department of Energy, Office of Science, Office of Basic Energy Sciences, under Contract No. DE-AC02-06CH11357. AFM, MAM, SM and DM acknowledge the support from the U. S. Department of Energy, Office of Science, Basic Energy Sciences, Materials Sciences and Engineering Division. Work performed at the National High Magnetic Field Laboratory is supported by National Science Foundation (NSF)
Cooperative Agreement No. DMR-1157490, the State of Florida, and the U.S. Department of Energy (DOE NNSA
DE-NA0001979).

This manuscript has been authored by UT-Battelle, LLC under Contract No. DE-AC05-00OR22725 with the U.S. Department of Energy. The United States Government retains and the publisher, by accepting the article for publication, acknowledges that the United States Government retains a non-exclusive, paid-up, irrevocable, world-wide license to publish or reproduce the published form of this manuscript, or allow others to do so, for United States Government purposes. The Department of Energy will provide public access to these results of federally sponsored research in accordance with the DOE Public Access Plan (http://energy.gov/downloads/doe-public-access-plan).

\end{acknowledgments}

%


\pagebreak
\setcounter{equation}{0}
\setcounter{figure}{0}
\setcounter{table}{0}
\setcounter{page}{1}
\makeatletter
\renewcommand{\theequation}{S\arabic{equation}}
\renewcommand{\thefigure}{S\arabic{figure}}
\renewcommand{\bibnumfmt}[1]{[S#1]}
\renewcommand{\citenumfont}[1]{S#1}

\renewcommand{\vec}[1]{\bm{\mathbf{#1}}}
\newcommand{\mat}[1]{\bm{\mathbf{#1}}}
\newcommand{\trp}[1]{{#1}^{\intercal}}
\newcommand{\vhat}[1]{\vec{\hat{#1}}}
\newcommand{\tr}{\textrm{tr}}
\newcommand{\hc}{\textrm{h.c.}}
\newcommand{\h}[1]{{#1}^{\dagger}} 
\newcommand{\cc}[1]{{#1}^{*}}
\newcommand{\cb}[1]{\bar{#1}}

\onecolumngrid

\begin{center}
\vspace{0.2 in}
\textbf{\large Supplemental materials for ``Candidate Elastic Quantum Critical Point in LaCu$_{6-x}$Au$_x$"}
\end{center}
\vspace{0.2 in}
\twocolumngrid
\section{Neutron and x-ray Diffraction}

\begin{table*}
\caption{Room temperature structural parameters of  a) LaCu$_6$ and b) LaCu$_{5.7}$Au$_{0.3}$. The parameters are obtained from the fits to the diffraction patterns shown in Fig. \ref{syn_diff}.}
\centering

a) {\bf LaCu$_6$}, space group = $P2_1/c$\\ 
$a_m$ = 5.145643(2) $\mathrm{\AA}$, $b_m$ = 10.20915(4) $\mathrm{\AA}$, $c_m$ = 8.14498(3) $\mathrm{\AA}$, $\beta_m$ = 91.485(1)\\
$R_P$ = 13.4, $~R_{wp}$ = 15.3, $~\chi^2$ =  2.9 \\

\begin{tabular}{lllllll}
    \hline
    \hline
   Element\hspace{30pt}	&	Wyck. \hspace{30pt}	&	$x/a$\hspace{40pt}				&	$y/b$ \hspace{40pt}	&	$z/c$	\hspace{40pt} & B \hspace{40pt}			&	Occ.	\\
    \hline
La1 & $4c$ & 0.2589(8)   & 0.4356(3)  & 0.2612(4)  & 0.69(4)  & 1\\
Cu1 & $4c$ & 0.0082(12)  & 0.1884(6)  & 0.4377(7)  & 0.61(12) & 1 \\
Cu2 & $4c$ & 0.2419(13)  & 0.1411(6)  & 0.1474(8)  & 0.82(12) & 1 \\
Cu3 & $4c$ & 0.2467(13)  & 0.7544(6)  & 0.8170(8)  & 0.83(12) & 1 \\
Cu4 & $4c$ & 0.2271(12)  & 0.6013(6)  & 0.5616(8)  & 0.82(12) & 1 \\
Cu5 & $4c$ & 0.2514(13)  & 0.5148(6)  & 0.8995(7)  & 0.49(11) & 1 \\
Cu6 & $4c$ & 0.4968(12)  & 0.1926(6)  & 0.4330(7)  & 0.62(11) & 1 \\
\hline
\end{tabular}

\vspace{5 pt}
b) {\bf LaCu$_{5.7}$Au$_{0.3}$}, space group = $Pnma$\\
$a$ = 8.19617(6) $\mathrm{\AA}$, $b$ = 5.13350(4) $\mathrm{\AA}$, $c$ = 10.27211(2) $\mathrm{\AA}$\\
$R_P$ = 13.1, $~R_{wp}$ = 14.7, $~\chi^2$ =  2.46 \\
\begin{tabular}{lllllll}
    \hline
    \hline
   Element\hspace{30pt}	&	Wyck. \hspace{30pt}	&	$x/a$\hspace{40pt}				&	$y/b$ \hspace{40pt}	&	$z/c$	\hspace{40pt} & B \hspace{40pt}			&	Occ.	\\
    \hline
La1 &   $4c$ &         0.2611(4)  & 0.2500            & 0.4360(3)  & 1.05(4)  & 1 \\
Cu1 &   $4c$ &        0.4361(5)  & 0.0029(10)        & 0.1897(5)  & 1.11(7)  & 1 \\
Cu2 &   $4c$ &        0.1462(4)  & 0.2500            & 0.1391(3)  & 0.87(8)  & 0.68(1)  \\
Au2 &   $4c$ &        0.1462(4)  & 0.2500            & 0.1391(3)  & 0.87(8)  & 0.32(1) \\
Cu3 &   $4c$ &        0.8180(8)  & 0.2500            & 0.7526(5)  & 1.04(11) & 1  \\
Cu4 &   $4c$ &        0.5614(9)  & 0.2500            & 0.6011(6)  & 1.18(11) & 1  \\
Cu5 &   $4c$ &        0.9040(7)  & 0.2500            & 0.5162(6)  & 1.09(12) & 1  \\
\hline
\end{tabular}
\label{syn_ref}
\end{table*}

The phase purity of polycrystalline \lacu samples was checked by laboratory x-ray diffraction measurements at room temperature. The diffraction patterns in \lacu corresponded to either the orthorhombic ($Pnma$) or the monoclinic ($P2_1/c$) phase depending on composition. For $x$ = 0.15, 0.2, 0.25, additional x-ray diffraction measurements were performed on a PANalytical X'Pert Pro MPD powder x-ray diffractometer using Cu$K_{\alpha,1}$ radiation ($\lambda~=1.5406~\mathrm{\AA}$). Full diffraction patterns were collected at room temperature and 20 K, followed by scans around few selected Bragg peaks in 10 K steps to determine the structural transition temperature, $T_S$.

\begin{figure}[h!]
\includegraphics[width=  0.45\textwidth]{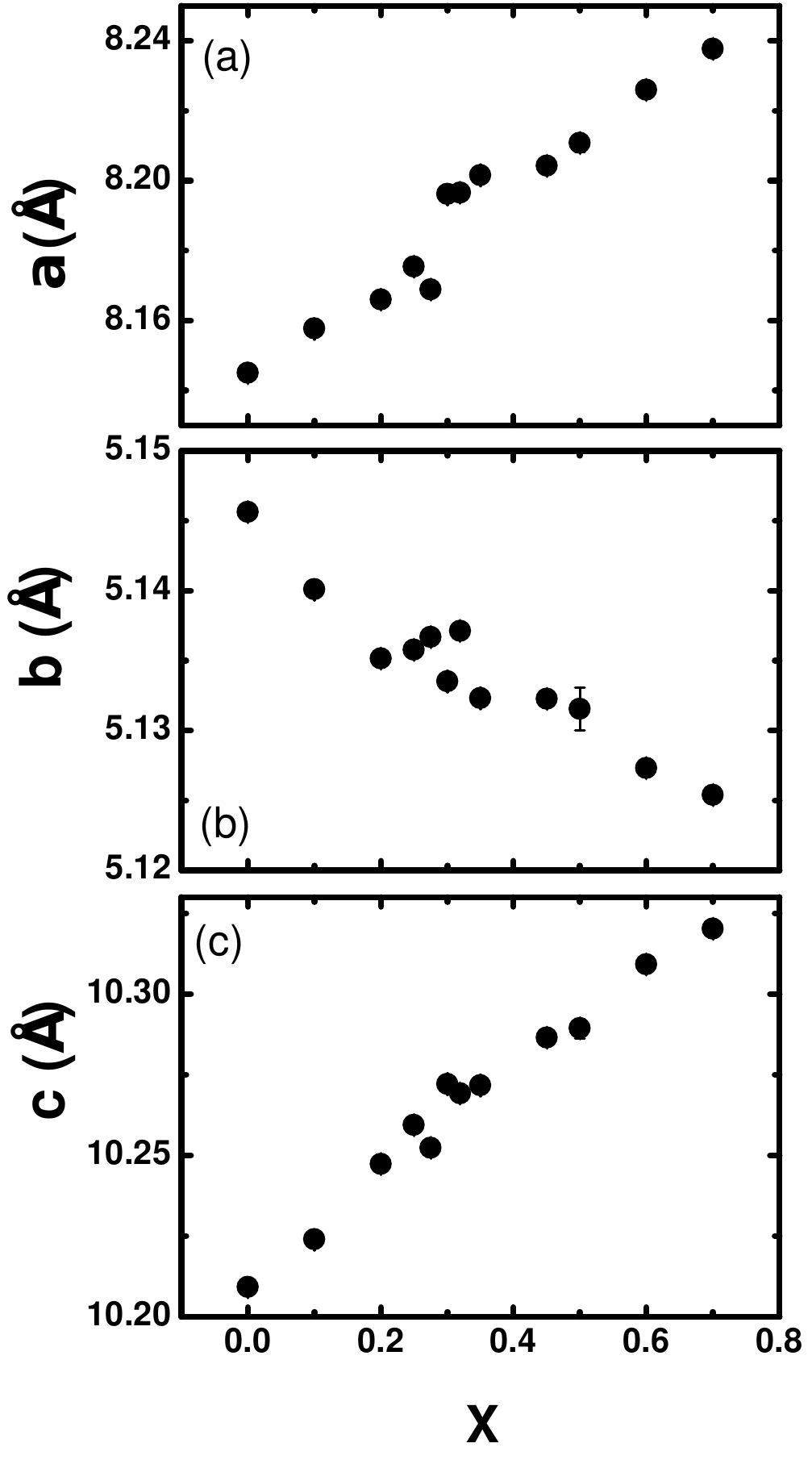}
\caption{Orthorhombic lattice parameters (a) $a$, (b) $b$, (c) $c$ in \lacu~ at room temperature. The lattice parameters were obtained from the Rietveld refinements of laboratory x-ray diffraction measurements. The compositions $x$ = 0 and 0.1 were refined in monoclinic structure (space group: $P2_1/c$). All other compositions ($x \geq$ 0.2) were refined to the orthorhombic phase (space group: $Pnma$). 
 }
\label{RT_parm}
\end{figure}

Neutron diffraction measurements were carried out on polycrystalline samples (LaCu$_{6-x}$Au$_x$, $x$ = 0.125, 0.265, 0.275, 0.3) with the HB-2A powder diffractometer at the High Flux Isotope Reactor (HFIR) of Oak Ridge National Laboratory. Neutrons of wavelength 1.54 ${\mathrm{\AA}}$ were used with collimations of $12^{\prime}-21^{\prime}-6^{\prime}$ before the monochromator, sample and detector, respectively. Samples of mass $\approx$ 5 g were finely powdered inside a glove box and held in a vanadium can with helium exchange gas to facilitate thermal conduction. The can was loaded in a closed cycle refrigerator system. Diffraction measurements were carried out at different temperatures near and below $T_S$. The structural parameters were extracted from Rietveld analyses \cite{Rietveld} of the diffraction patterns using the FullProf software suite \cite{rodriguez1990fullprof}.

\begin{figure}[b]
\includegraphics [width=  0.45\textwidth]{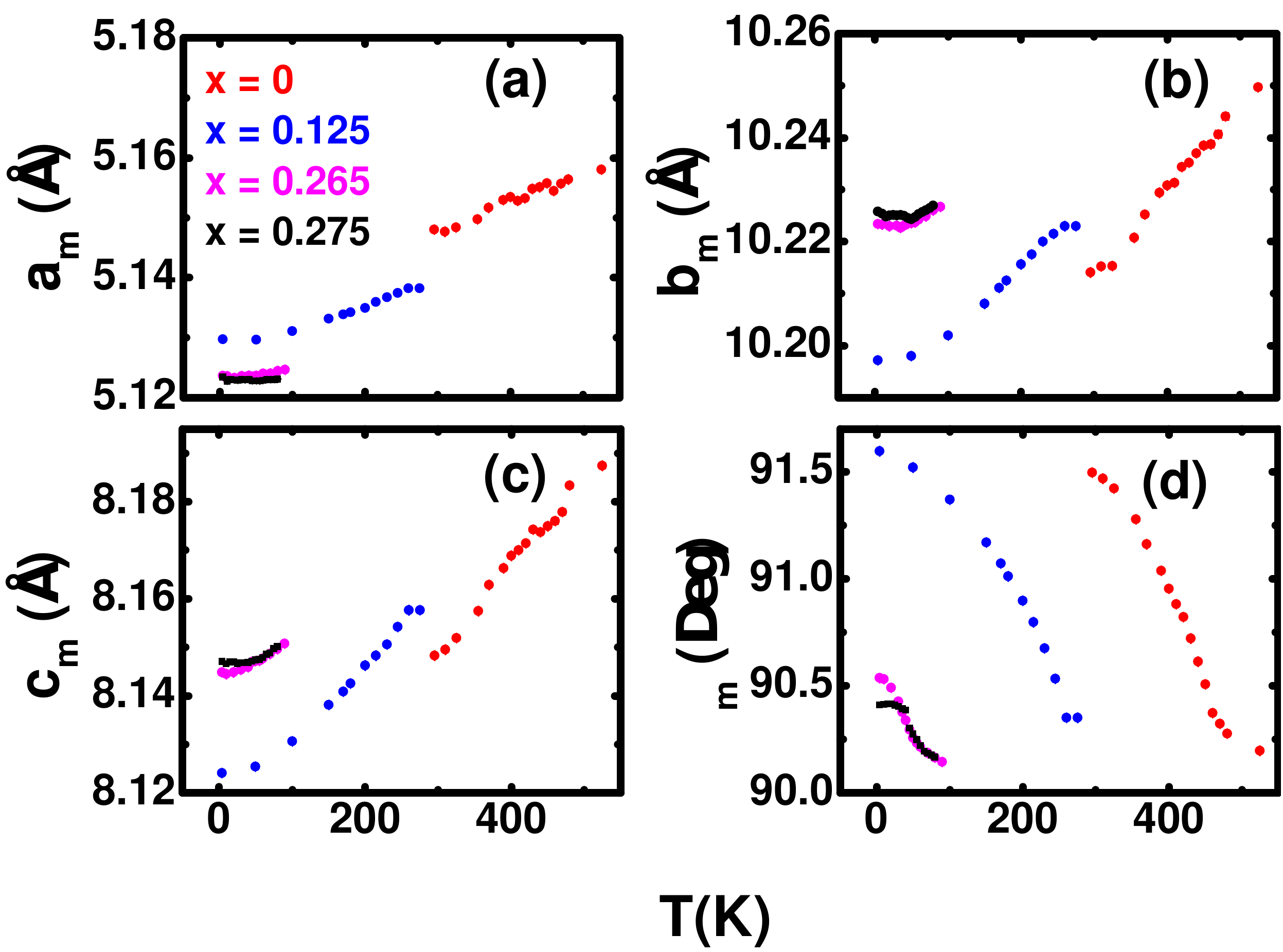}
\caption{Temperature dependence of the monoclinic lattice parameters (a) $a_m$, (b) $b_m$, (c) $c_m$, and (d) $\beta_m$ in \lacu.}
\label{parm}
\end{figure}

Figure \ref{RT_parm} displays the compositional variation of the orthorhombic lattice parameters in \lacu at room temperature. With an increase in Au-composition, the $b$-parameter decreases while $a$ and $c$-parameter show a linear increase with $x$. The temperature dependence of the lattice parameters in \lacu~ (for $x~=~0,~0.125,~0.265,~0.275$) is shown in Fig. \ref{parm}. In agreement with the continuous nature of the phase transition, the lattice parameters change continuously with temperature and Au-composition.

The structural phase transition takes place with a continuous evolution of shear strain. For the orthorhombic-monoclinic phase transition, the shear-component of the strain tensor can be written as \cite{carpenter1998_2},

\begin{equation}
    e_{13}=\frac{1}{2}\Bigg(\frac{c}{c_o} \cos\beta_m \Bigg)
\end{equation}

\begin{figure}[b]
\includegraphics[width= 0.45\textwidth]{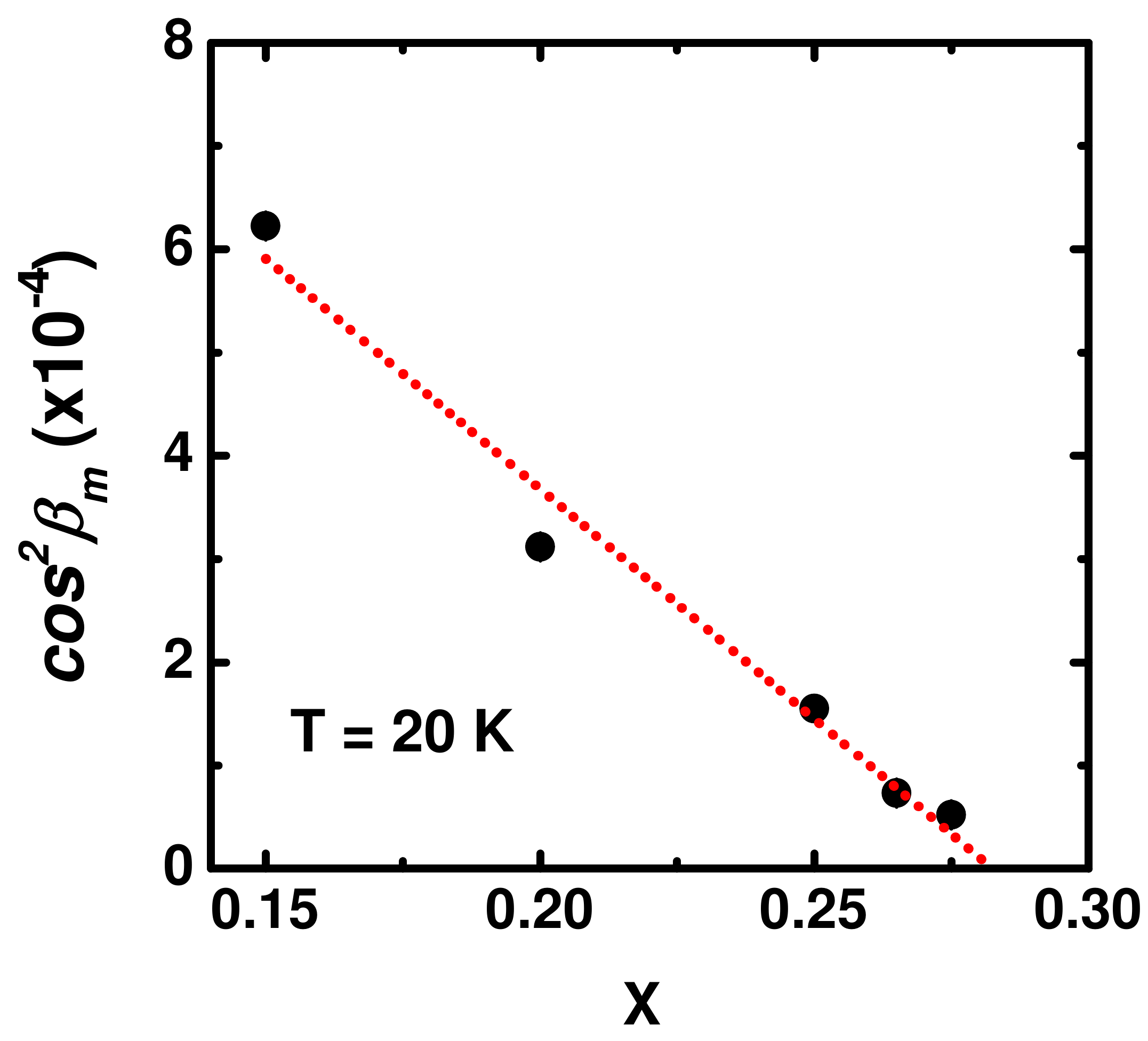}
\caption{The variation of $\cos^2\beta_m$ in \lacu~ with
Au-composition at 20 K. An extrapolation (dotted red line) of $\cos^2\beta_m$ with Au-composition indicates that the suppression of the monoclinic ground state occurs at $x= 0.28$. }
\label{OPphase}
\end{figure}

\begin{figure*}
\centering
\includegraphics[width=0.9\textwidth]{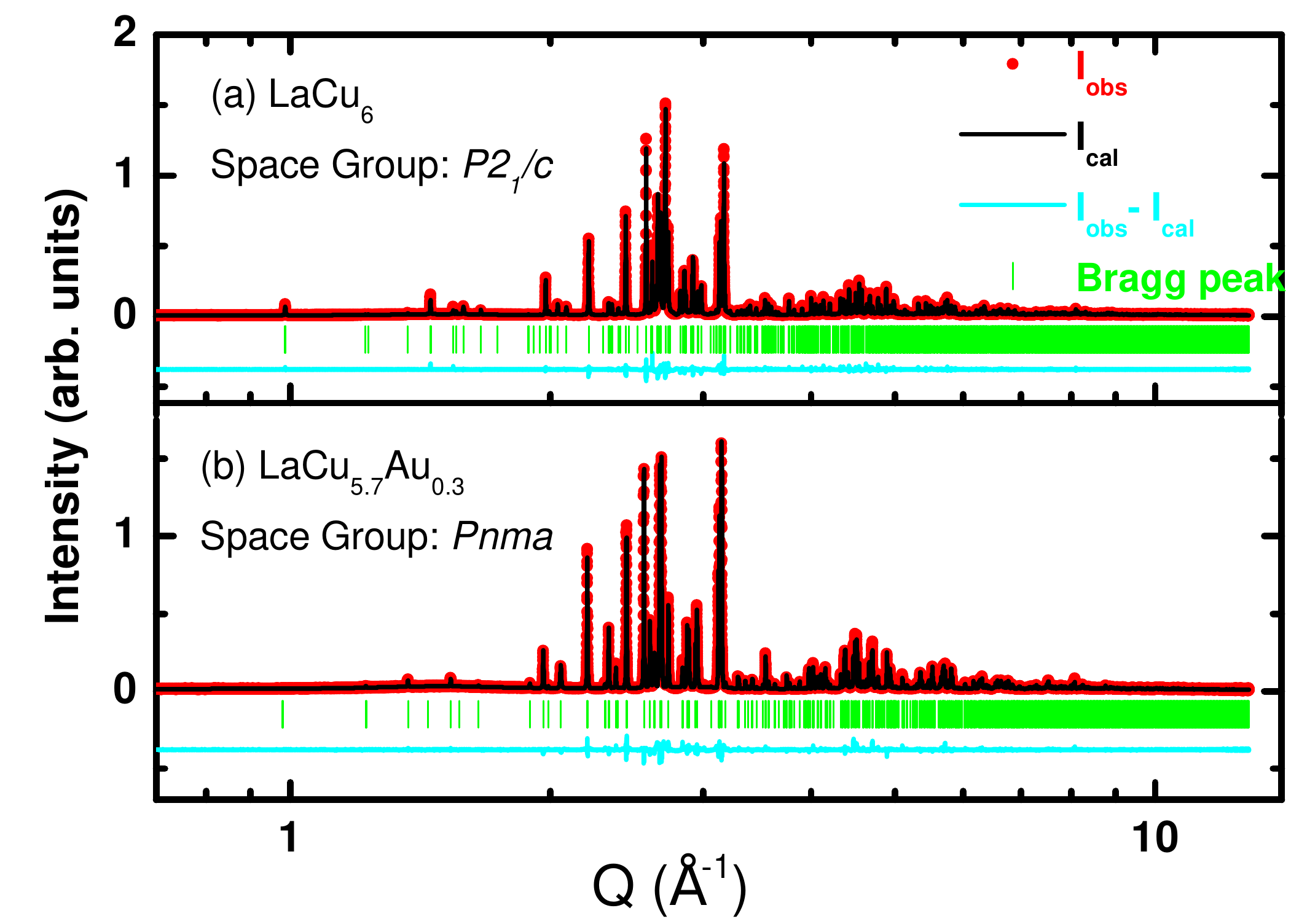}
\caption{Synchrotron x-ray diffraction pattern of (a) LaCu$_6$, (b) LaCu$_{5.7}$Au$_{0.3}$ at 300 K. The fit (black) to the diffraction pattern (red) is obtained with the Rietveld analysis. The vertical green bars represent structural Bragg peaks. $\mathrm{I_{calc} - I_{obs}}$ is offset for clarity. Note: x-axis is in logarithmic scale. }
\label{syn_diff}
\end{figure*}

\begin{figure}[t]
\includegraphics[width=  0.45\textwidth]{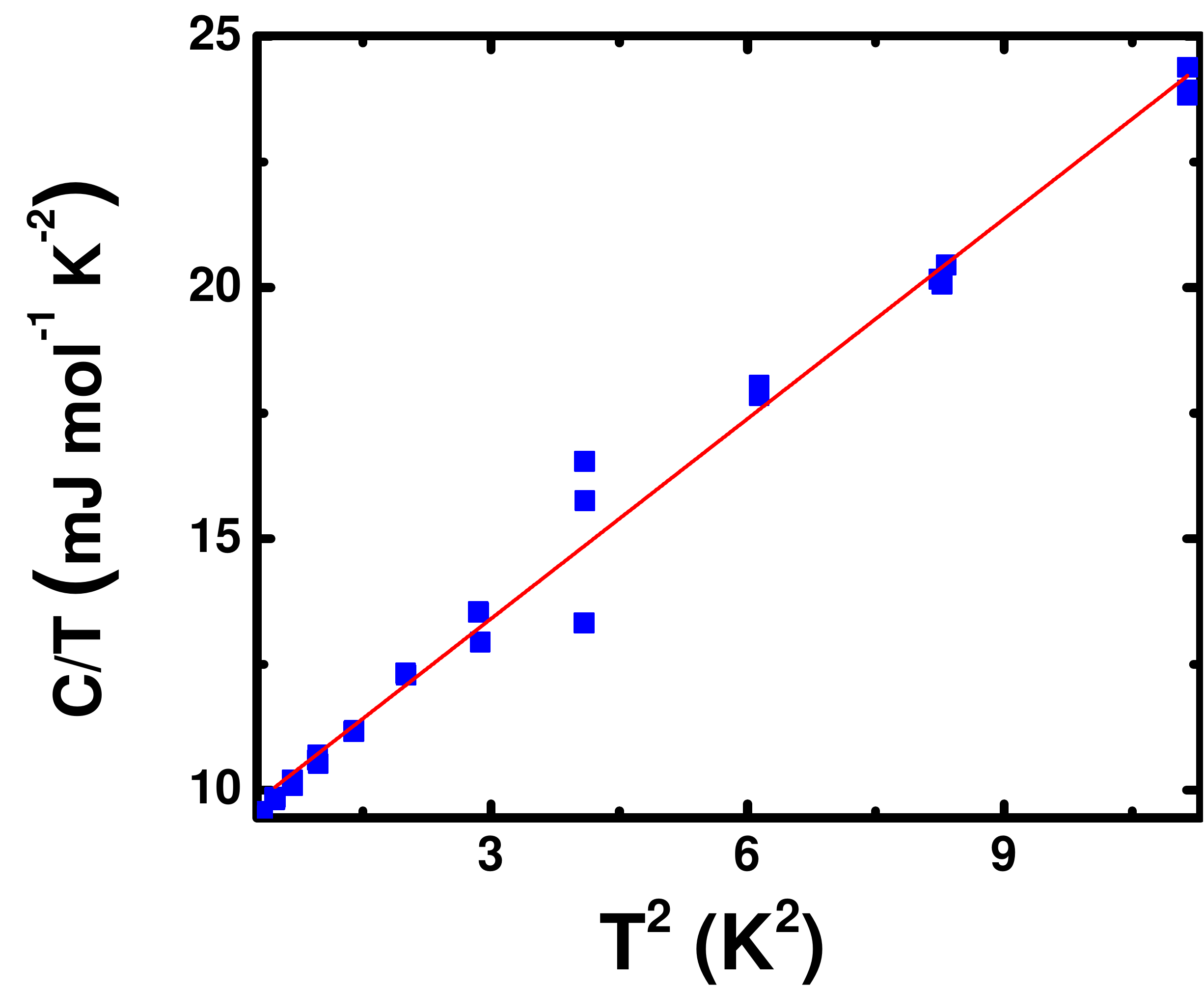}
\caption{Heat capacity measurement of LaCu$_{5.7}$Au$_{0.3}$ showing Debye-$T^3$ behavior at low temperatures (0.7 K $\geq$ T $\geq$ 3.5 K). The straight line (red) is the fit of the equation $C/T = \gamma  + \beta T^2$ to the data.}
\label{heat_supp}
\end{figure}

\begin{figure}[b]
\includegraphics[width=  0.45\textwidth]{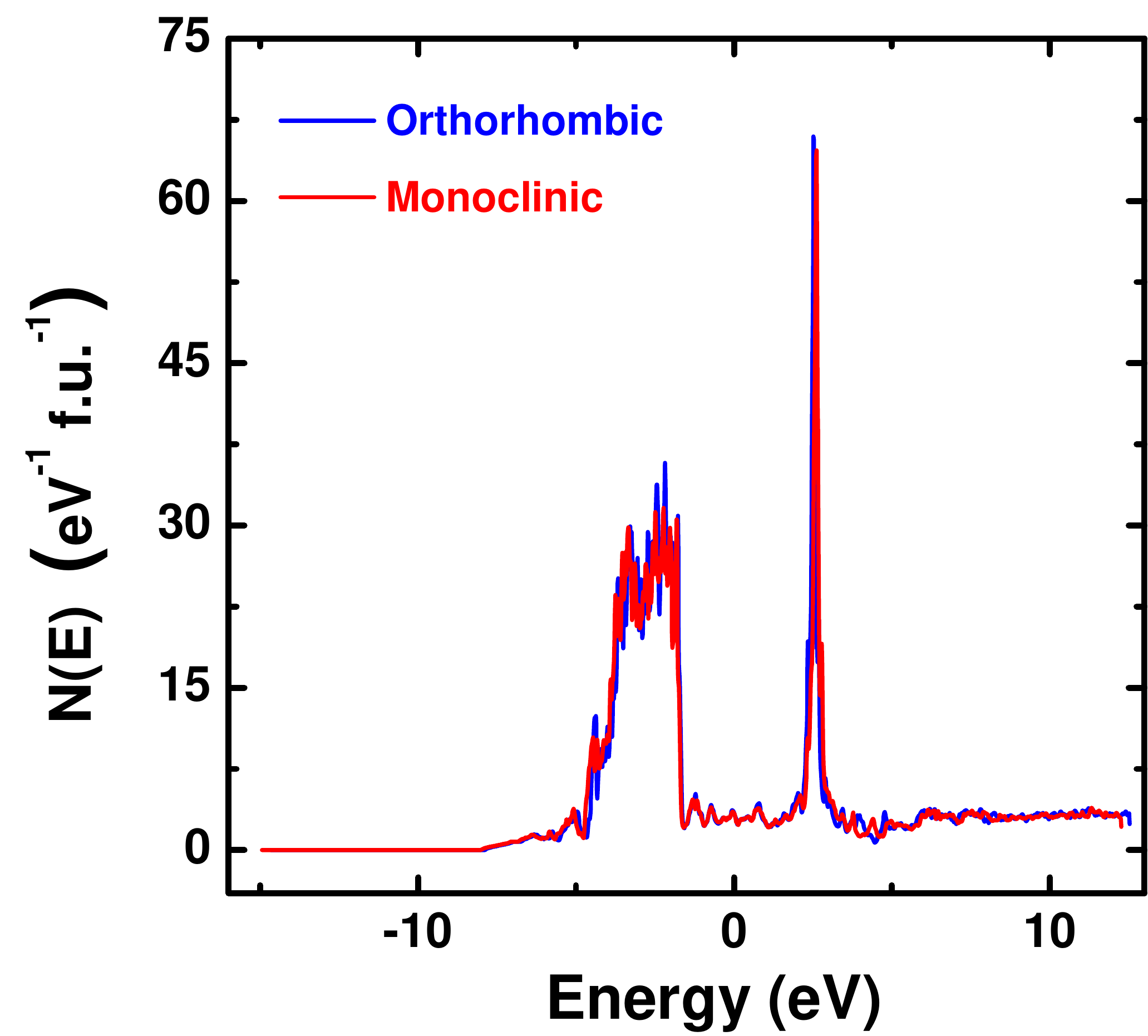}
\caption{Electronic density of states of LaCu$_6$ in orthorhombic and monoclinic phases. }
\label{DOS}
\end{figure}

where, $c_o$ is the extrapolation of lattice parameter in the orthorhombic structure into the monoclinic structure. The shear strain is zero in the orthorhombic phase. 

For the estimation of the transition temperature, $\cos^2{\beta_m}$ is linearly extrapolated to zero value for a particular composition (Fig. 1b of main paper). At 20 K, a similar extrapolation of $\cos^2{\beta_m}$ as a function of Au-composition, rather than temperature, yields $x_c$ = 0.28 which is close to the value $x_c$ = 0.3 determined from directly extrapolating $T_S$.

The $d-$spacing corresponding to a reflection in orthorhombic and monoclinic geometry is defined as, 
\begin{equation}
    \frac{1}{d_{orthorhombic}^2} = \bigg(\frac{h^2}{a^2}+\frac{k^2}{b^2}+\frac{l^2}{c^2}\bigg)
\end{equation}
\begin{equation}
    \frac{1}{d_{monoclinic}^2} =\frac{1}{\sin^2{\beta_m}} \bigg(\frac{h^2}{a_m^2}+\frac{k^2}{b_m^2}+\frac{l^2}{c_m^2}-\frac{2hl\cos{\beta_m}}{a_mc_m}\bigg)
    \label{mono}
\end{equation}

In the orthorhombic structure, the $d-$spacing of each member of a Bijvoet pairs is equal. However, in the monoclinic structure, for $h~\neq~0$ and $l~\neq~0$, the $d-$spacing for two Bijvoet pairs can be different. Consequently, in a powder diffraction measurement, a structural Bragg peak that can be indexed by ($hkl$) in the orthorhombic system separates into a pair of peaks which correspond to ($kl$-$h$) and (-$klh$) in the monoclinic phase \footnote{Note: The cyclic change of lattice parameters when going from the orthorhombic to monoclinic notations.}. The temperature at which the peaks begin to separate is characterized as the transition temperature ($T_S$), as illustrated in Fig. (1)a of the main text.

\begin{table}
\centering
\caption{The goodness of fit parameters at different stages of Rietveld refinement of LaCu$_{5.7}$Au$_{0.3}$. The best set of parameters (shown in bold)  was obtained when the Au atoms were constrained to the Cu2 site.}
\label{ref}
\begin{tabular}{llll}
\hline
    \hline
\begin{tabular}[c]{@{}l@{}}Au distributed in \hspace{30pt}\end{tabular} & $R_p$\hspace{25pt}            & $R_{wp}$ \hspace{25pt}          & $\chi^2$          \\
\hline
Cu1-site        & 32.7\%          & 39.6\%          & 17.95        \\
\textbf{Cu2-site}                                             & \textbf{13.4\%} & \textbf{15.3\%} & \textbf{2.9} \\
Cu3-site         & 28.6\%          & 37.5\%          & 16.06         \\
Cu4-site         & 26.9\%          & 68.5\%          & 53.71         \\
Cu5-site         & 28.3\%          & 34.5\%          & 13.61        \\
Equally distributed   \hspace{30pt}   & 22.7\%          & 25.7\%          & 7.54          \\ 
\hline 
\end{tabular}
\end{table}

High resolution synchrotron x-ray diffraction measurements were performed on two compositions, LaCu$_6$ and LaCu$_{5.7}$Au$_{0.3}$, for a more detailed analysis of the orthorhombic and monoclinic structures. The measurements were carried out at beamline 11-BM of the Advanced Photon Source (APS) at Argonne National Laboratory (ANL) using synchrotron x-rays of wavelength $\lambda = 0.413 ~\mathrm{\AA}$. The samples were mixed with SiO$_2$ powder in a molar ratio of 1:2 (for LaCu$_6$) and  1:3 (for LaCu$_{5.7}$Au$_{0.3}$) to minimize x-ray absorption. The finely ground mixtures were held in kapton tubes of diameter 0.8 mm for the measurements.

Figure \ref{syn_diff} displays room temperature synchrotron x-ray diffraction patterns with the fits obtained from Rietveld refinements. In agreement with the compositional dependence displayed in the phase diagram (Fig. 2 of the main text), high resolution synchrotron diffraction measurements at room temperature confirm that the crystal structures of LaCu$_6$ and LaCu$_{5.7}$Au$_{0.3}$ are monoclinic ($P2_1/c$) and orthorhombic ($Pnma$), respectively. No impurity peaks were observed in synchrotron x-ray diffraction measurements, consistent with phase pure samples. The structural and goodness of fit parameters determined from the Rietveld analysis are presented in Table \ref{syn_ref}.

The Rietveld refinement of LaCu$_{5.7}$Au$_{0.3}$ was carried out in various steps. In each stage, Au atoms were either placed in one of the five different Cu sites or equally distributed among them. A comparison between calculated and observed intensities at different stages of the refinements indicates that the Au atoms in LaCu$_{5.7}$Au$_{0.3}$ exclusively occupy the Cu2 site located at ~($x/a~=~0.146,~y/b~=~0.25,~z/c~=~0.139$). As presented in the Table \ref{ref}, the best set of goodness of fit parameters, $R_p$, $R_{wp}$ and $\chi^2$, are obtained when the Au atoms are constrained to Cu2 site. Furthermore, the refined occupancy of Au on Cu2 site is 0.31(1), which coincides well with the expected value 0.3 (Table \ref{syn_ref}(b)).

\section{ Heat capacity}

Heat capacity measurements of \lacu~($x$= 0, 0.1, 0.2, 0.25, 0.27, 0.3, 0.32, 0.35, 0.4, 0.5, 0.6, 0.7) above 2 K were performed in a Quantum Design PPMS system. For all compositions studied, the low temperature heat capacity was modelled by the relation, $C = \gamma T  + \beta T^3$ \cite{kittel2}, where $\gamma$ and $\beta$ capture the electronic and phonon contributions to the heat capacity, respectively. 

To check for an enhanced superconducting transition temperature at the critical concentration over the 0.16 K T$_c$ in LaCu$_6$ \cite{supercond}, an additional heat capacity measurement of LaCu$_{5.7}$Au$_{0.3}$ was performed down to lower temperatures using a PPMS with $^3$He insert. No superconducting transition was observed above 0.7 K. The heat capacity data are consistent with the Debye model, as shown in Fig. \ref{heat_supp}.

\begin{figure}[t]
\includegraphics[width=  0.45\textwidth]{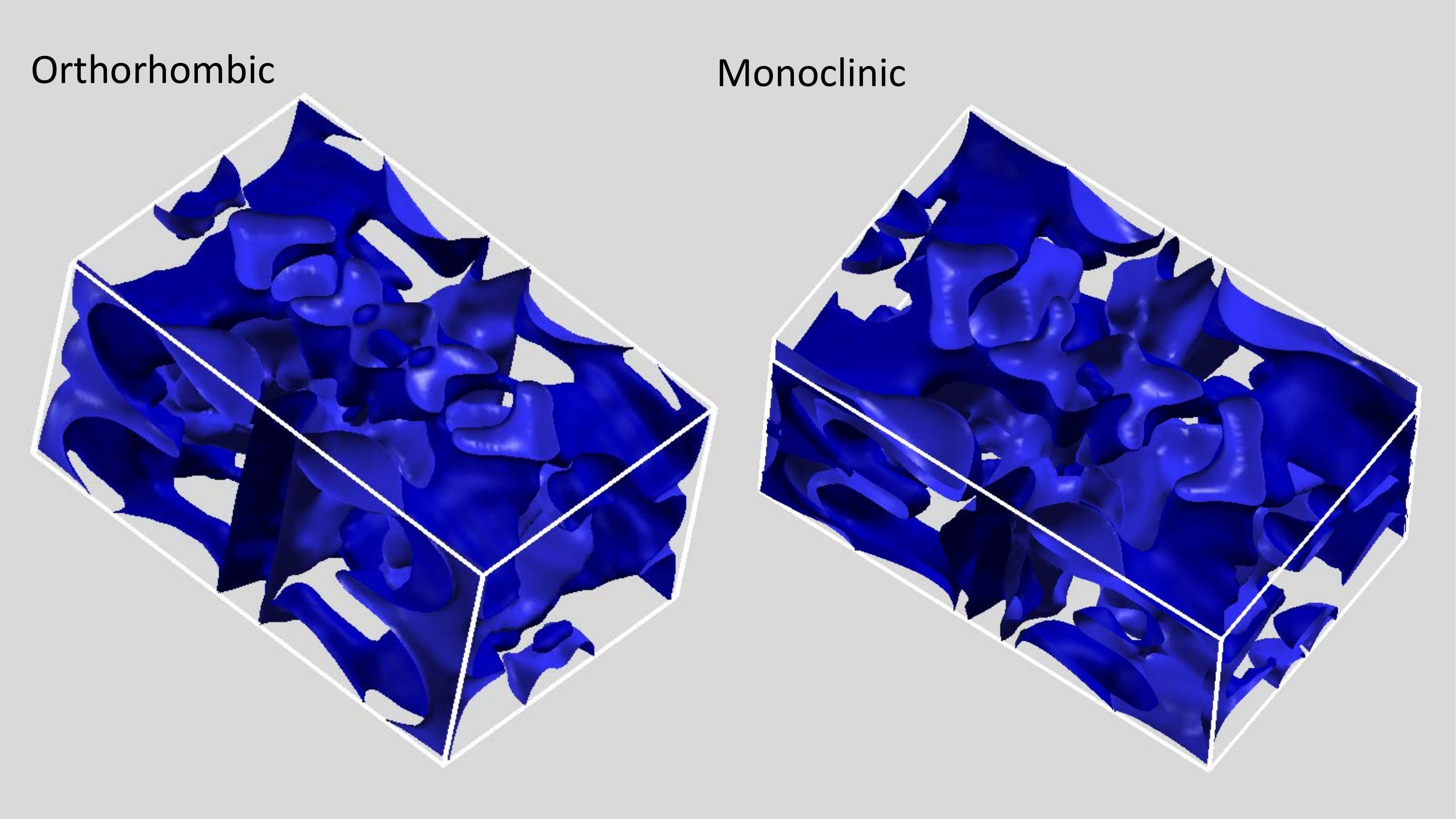}
\caption{Fermi surfaces of LaCu$_6$ in orthorhombic and monoclinic phases. }
\label{fermi-both}
\end{figure}

\section{Density Functional Theory Calculations}
Figure \ref{DOS} shows the electronic density of states for orthorhombic and monoclinic phases of LaCu$_6$. The Fermi surfaces of \lacu are shown in Fig. \ref{fermi-both}.
%


\begin{thebibliography}{55}
\bibitem [{not()}]{note1}%
  \BibitemOpen
  \href@noop {} {\bibinfo  {journal} {Note that the statistical error bar on
  this extrapolation is very small, $\pm$0.004. A more realistic upper bound
  for the error on $x_c$ is given by the interval between the final two
  compositions which is 0.025}\ }\BibitemShut {NoStop}%
\end{thebibliography}
\end{document}